\documentclass[aps, pra, twocolumn, superscriptaddress]{revtex4-2}

\usepackage[pdftex]{graphicx}
\usepackage{amsmath}
\usepackage{amssymb}
\usepackage{braket}
\usepackage{siunitx}
\usepackage{ragged2e}
\setcitestyle{numbers}
\usepackage{subcaption}
\usepackage{enumitem}
\usepackage{placeins}

\usepackage[dvipsnames]{xcolor}
\usepackage{hyperref}
\hypersetup{
  colorlinks   = true, 
  urlcolor     = MidnightBlue, 
  linkcolor    = MidnightBlue, 
  citecolor   = Maroon 
}

\renewcommand{\arraystretch}{2.0}

\begin{document}

\title{Demonstration of measurement-free universal fault-tolerant quantum computation} 

\author{Friederike Butt}
\altaffiliation{These authors contributed equally}
\affiliation{Institute for Quantum Information, RWTH Aachen University, Aachen, Germany}
\affiliation{Institute for Theoretical Nanoelectronics (PGI-2), Forschungszentrum J\"{u}lich, J\"{u}lich, Germany}

\author{Ivan Pogorelov}
\altaffiliation{These authors contributed equally}
\affiliation{Universit\"{a}t Innsbruck, Institut f\"{u}r Experimentalphysik, Innsbruck, Austria}

\author{Robert Freund}
\affiliation{Universit\"{a}t Innsbruck, Institut f\"{u}r Experimentalphysik, Innsbruck, Austria}

\author{Alex Steiner}
\affiliation{Universit\"{a}t Innsbruck, Institut f\"{u}r Experimentalphysik, Innsbruck, Austria}

\author{Marcel Meyer}
\affiliation{Universit\"{a}t Innsbruck, Institut f\"{u}r Experimentalphysik, Innsbruck, Austria}

\author{Thomas Monz}
\affiliation{Universit\"{a}t Innsbruck, Institut f\"{u}r Experimentalphysik, Innsbruck, Austria}
\affiliation{Alpine Quantum Technologies GmbH, Innsbruck, Austria}

\author{Markus M\"{u}ller}
\email[Email to ]{markus.mueller@fz-juelich.de}
\affiliation{Institute for Quantum Information, RWTH Aachen University, Aachen, Germany}
\affiliation{Institute for Theoretical Nanoelectronics (PGI-2), Forschungszentrum J\"{u}lich, J\"{u}lich, Germany}

\date{\today}

\begin{abstract}

The ability to perform quantum error correction (QEC) and robust gate operations on encoded qubits opens the door to demonstrations of quantum algorithms. Contemporary QEC schemes typically require mid-circuit measurements with feed-forward control, which are challenging for qubit control, often slow, and susceptible to relatively high error rates. 
In this work, we propose and experimentally demonstrate a universal toolbox of fault-tolerant logical operations without mid-circuit measurements on a trapped-ion quantum processor. 
We present modular logical state teleportation between two four-qubit error-detecting codes without measurements during algorithm execution. 
Moreover, we realize a fault-tolerant universal gate set on an eight-qubit error-detecting code hosting three logical qubits, based on state injection, which can be executed by coherent gate operations only. 
We apply this toolbox to experimentally realize Grover's quantum search algorithm fault-tolerantly on three logical qubits encoded in eight physical qubits, 
with the implementation displaying  clear identification of the desired solution states. 
Our work demonstrates the practical feasibility and provides first steps into the largely unexplored direction of measurement-free quantum computation. 

\end{abstract}

\maketitle

\subsection{Introduction}

\begin{figure*}[!tb]
	\centering
	\includegraphics[width=180mm]{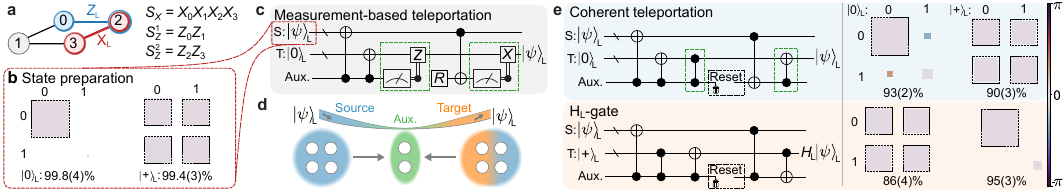}
 \caption{\justifying \textbf{Measurement-free logical state teleportation with the $[[4, 1, 2]]$-code. } \textbf{a,} Stabilizers $S_Z$, $S_X$ and logical operators of the $[[4, 1, 2]]$-code. \textbf{b,} Experimental logical quantum state tomography for FT logical state initialization. The black dashed boxes correspond to ideal values in a fault-free case. \textbf{c,} High-level circuit for \textit{measurement-based} modular logical teleportation. The source (S) and target (T) code blocks are merged by measuring the joint logical $X_\mathrm{L}^\mathrm{S} X_\mathrm{L}^\mathrm{T}$-operator via an auxiliary register (Aux.) and applying a $Z$-type feedback operation based on the measurement outcome (first green box). The two blocks are then split again by measuring $Z_\mathrm{L}^\mathrm{S}$ and applying a $X$-type operation to the target register conditioned on the measurement outcome (second green box). \textbf{d,} Schematic illustration of modular state teleportation, where the source and target registers are never directly coupled to one another, but only interact via an auxiliary quantum register (Aux).  
 \textbf{e,} High-level circuits for \textit{measurement-free} logical teleportation and experimental logical quantum state tomography. We replace the measurements and feed-forward operations with coherent feedback operations to teleport a state without mid-circuit measurements (blue). An additional $H_\mathrm{L}$ is applied to the target state using the circuit shown in orange. The reset operation can either be carried out explicitly by physically resetting the auxiliary qubits and reusing them afterwards, or implemented by replacing them with fresh qubits. 
 }
	\label{fig:teleportation}
\end{figure*}

The practical implementation of quantum algorithms depends on their resilience to errors, alongside the ability to perform arbitrary quantum operations. Quantum error correction (QEC) enables the detection and correction of errors arising during computation by encoding information across multiple physical qubits~\cite{preskill2018quantum, gottesman1997stabilizer, Nielsen_and_Chuang}. Computations on these encoded qubits can be realized through a discrete, universal set of gates~\cite{Nielsen_and_Chuang}. These operations have to be implemented in a robust, fault-tolerant (FT) fashion, meaning that local faults in the underlying gate operations do not proliferate uncontrollably across the logical qubits~\cite{knill1998resilient}. However, no QEC code intrinsically supports a full, inherently FT universal gate set~\cite{eastin2009restrictions}. Completing this FT universal gate set is a key challenge for realizing a potential advantage beyond the reach of algorithms that can be efficiently simulated classically. %for any practical application of quantum computing. 
Recent experiments have demonstrated QEC cycles on trapped-ion quantum processors~\cite{ryan2021realization, postler2023demonstration, huang2023comparing, reichardt2024demonstration, nguyen2021demonstration}, superconducting architectures~\cite{acharya2024quantum, lacroix2024scaling, krinner2022realizing, zhao2022realization}, as well as neutral-atom platforms~\cite{bluvstein2024logical, bedalov2024fault}. 
FT universal gate sets have been realized on these platforms by means of code switching~\cite{pogorelov2025experimental, daguerre2025experimental}, where information is transferred between two codes with complementary sets of inherently FT gates, as well as magic-state injection~\cite{dasu2025breaking, ryan2024high, postler2022demonstration, lacroix2024scaling}, which requires high-fidelity magic states as a resource~\cite{rodriguez2024experimental, gupta2024encoding}. 
These advancements in the practical and scalable implementations of logical qubits enabled the execution of first, small quantum algorithms run on encoded qubits, such as the Bernstein-Vazirani algorithm~\cite{bernstein1993quantum, reichardt2024logical}, one-bit addition~\cite{wang2024fault} or the quantum Fourier transform on three logical qubits~\cite{mayer2024benchmarking}.  

Many practical protocols rely on measurements during algorithm execution and feed-forward operations conditioned on these measurement outcomes, which is experimentally demanding on many hardware platforms and limits their success probability: In both atomic and superconducting quantum processors, 
measurements remain orders of magnitude slower than typical gate times, which poses speed limitations and results in decoherence of idling qubits during measurements. 
Moreover, fluorescence read-out in atomic setups requires additional cooling during and after measurements, as atoms are heated during this process~\cite{pogorelov2025experimental, graham2023midcircuit, singh2023mid, ryan2024high, moses2023race}. 

Following early works~\cite{paz2010fault, gottesman2016surviving}, recent theoretical works have proposed  \textit{practical measurement-free} protocols for logical state preparation~\cite{goto2023measurement}, rounds of QEC~\cite{perlin2023fault, heussen2024measurement, veroni2024optimized} and the implementation of a FT universal gate set~\cite{butt2024measurement, veroni2024universal, brechtelsbauer2025measurement}. In these protocols, stabilizer information is transferred onto auxiliary qubits, allowing decoding and coherent feedback to be carried out within the quantum algorithm itself. This approach avoids the need for mid-circuit measurements or feed-forward operations entirely. At the end, auxiliary qubits are replaced or reset to be reused, effectively removing the entropy introduced by the noise. 

In this work, we develop and experimentally demonstrate a complete toolbox of logical operations needed for FT universal quantum computing on an ion-trap quantum processor, without mid-circuit measurements or feed-forward operations. First, we construct protocols for modular logical quantum state teleportation, such that different encoded blocks are never directly coupled to one another, which is a key desideratum for scaling up quantum computations to large numbers of logical qubits. We analyze the performance of these measurement-free protocols for different logical input states, accompanied by numerical simulations. We then complete a FT, measurement-free universal gate set for an eight-qubit error-detecting code by constructing and implementing circuits for a logical Hadamard-gate on an encoded qubit. 
Finally, we use this implementation as a building block for Grover's algorithm to search two elements out of eight, for the first time demonstrating a small-scale FT and measurement-free universal quantum algorithm. 

\subsection{Experimental setup}\label{sec:exp_setup}

The experimental data was obtained with a 16-qubit quantum computing device based on trapped ions \cite{pogorelov2021compact}. 
The chain of 16 $^{40}$\textrm{Ca}$^+$ ions is confined in a linear Paul trap.
The physical qubits are encoded in $\ket{0}=\ket{4 ^2\textrm{S}_{1/2}, m_J = -1/2}$ and $\ket{1}=\ket{3 ^2\textrm{D}_{5/2}, m_J = -1/2}$ Zeeman sub-levels. 
The state of each qubit can be manipulated individually by optically addressing the ions with \SI{729}{\nano\meter} laser light. Two-qubit gates are realized as a M{\o}lmer-S{\o}rensen (MS) interaction~\cite{sorensen2000entanglement}, providing all-to-all two-qubit-gate connectivity. Overall, the native gate set of the device includes arbitrary-angle rotation gates $R(\theta,\phi) = \exp(-i\frac{\theta}{2}[X \cos \phi + Y \sin \phi])$, `virtual' $Z$-gates $R_Z(\theta)=\exp(-i\frac{\theta}{2}Z)$, and maximally-entangling two-qubit gates $XX(\pi/2) = \exp(- i\frac{\pi}{4} X\otimes X)$. A description of the experimental setup can be found in \cite{pogorelov2021compact, postler2022demonstration, heussen2023strategies}. 

Our trapped-ion platform is capable of performing mid-circuit measurement operations, as was shown in \cite{postler2023demonstration}. However, such an operation, together with an additional feed-forward, represents a substantial experimental overhead in both sequence duration and infidelity. 
In our protocols, we do not need to perform these operations but require only resets of the quantum state of certain qubits, which is discussed further in Appendix~\ref{app:reset}. Instead of re-initializing physical qubits, one can also replace them with fresh physical qubits. In our experiments, we make use of the full 16-ion register and use fresh physical qubits whenever possible. 

\subsection{Logical state teleportation without mid-circuit measurements}\label{sec:MF_teleportation}

In this section, we discuss how to teleport a logical state between two four-qubit registers without mid-circuit measurements or feed-forward operations, and demonstrate this concept experimentally. 
We consider a $[[4, 1, 2]]$-code instance that encodes $k = 1$ logical qubit in $n = 4$ physical qubits and has distance $d = 2$, meaning that any single error can be detected. The stabilizers and logical operators defining the code are shown in Fig.~\ref{fig:teleportation}\textbf{a}. 

A standard approach for teleporting a state between registers is based on lattice surgery~\cite{horsman2012surface, gutierrez2019transversality}, which is illustrated in Fig.~\ref{fig:teleportation}\textbf{c}. First, two code blocks are merged by measuring the joint logical $X$-operator. Based on this measurement outcome, one applies a logical $Z$-operation to the target register. In a second step, the two blocks are split again by measuring the logical $Z$-operator of the source register and applying a conditional logical $X$-operation to the target register. The measurement-based approach has been realized experimentally on various platforms~\cite{zhang2025leveraging, erhard2021entangling, besedin2025realizing, ryan2024high, lacroix2024scaling}. 

Instead of performing measurements and conditional operations based on the measurement outcomes, we now map the respective operators to an auxiliary register and apply coherent feedback operations, as illustrated in Fig.~\ref{fig:teleportation}\textbf{e} in blue. In the first step, we couple both logical qubit registers to the auxiliary register by applying pairs of CNOT-gates to map the information about the joint logical operator $X_\mathrm{L}^\mathrm{S} X_\mathrm{L}^\mathrm{T}$ of the source (S) and the target (T) register to the auxiliary qubits. 
The conditional logical $Z$-operation can then be implemented coherently with a combination of C$Z$-gates that act on the auxiliary and target register, as shown in the green dashed box in Fig.~\ref{fig:teleportation}\textbf{e}. In the second step, we map the logical $Z_\mathrm{L}^\mathrm{S}$ to the auxiliary register and apply a coherent feedback with a combination of CNOT-gates. 
The scheme is made FT by repeating subroutines, i.e.~by mapping multiple stabilizer-equivalent logical operators onto auxiliary qubits, as discussed further in Apps.~\ref{app:anticpiated_teleportation} and~\ref{app:numerical_methods}. The explicit circuits can be found in App.~\ref{app:circuits}. 

We construct a similar protocol that enables the implementation of a logical $H_\mathrm{L}$-gate (shown in Fig.~\ref{fig:teleportation}\textbf{e} in orange) with the same resources as the bare teleportation protocol. We find this circuit by inserting a physical $H$-gate to the target qubit and propagating it back through the circuit, such that no $H$-gate has to be performed explicitly. This means that no $H_\mathrm{L}$-gate has to be applied to a $[[4, 1, 2]]$ instance when shifting this circuit to the logical level. 

We experimentally perform logical state tomography for three protocols: state initialization, logical state teleportation and the application of a logical $H_\mathrm{L}$-gate for logical input states $|0\rangle_\mathrm{L}$ and $|+\rangle_\mathrm{L}$, which is shown in Fig.~\ref{fig:teleportation}\textbf{b, e}. Further details on the measurement bases and number of shots can be found in Apps.~\ref{app:tomography} and~\ref{app:number_measurements}. 
We achieve fidelities of up to $93(2)\%$ for state teleportation and $95(3)\%$ for a $H_\mathrm{L}$-gate. The difference in fidelities for the two logical input states can be traced back to two sources. First, dephasing on idling qubits due to fluctuations in the magnetic fields introduces a strong bias towards $Z$-type errors. Furthermore, we measure the qubits in the $Z$-basis in the end and determine the logical value from this measurement, if the target state is a $|0\rangle_\mathrm{L}$-state. Based on these outcomes, we perform a classical round of error detection and postselect on the two $Z$-stabilizers of the $[[4, 1, 2]]$-code. When the target state is the $|+\rangle_\mathrm{L}$-state and we determine the logical $X$-value, we can only postselect on one $X$-stabilizer. By accepting fewer runs, we effectively also discard a fraction of runs where higher-weight errors lead to a failure, and fidelities increase the more we postselect. 

\begin{figure}[!t]
	\centering
	\includegraphics[width=86mm]{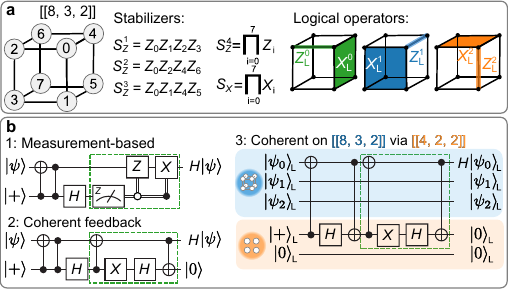}
 \caption{\justifying \textbf{FT logical operations on an $[[8, 3, 2]]$-code. } \textbf{a,} Definition of stabilizers and logical operators on the $[[8, 3, 2]]$-code~\cite{campbell2016smallest, wang2024fault}. \textbf{b,} The upper left circuit implements \textit{measurement-based} $H$-gate injection, where an auxiliary qubit is prepared in $|+\rangle$ and entangled with the data qubit in state $|\psi\rangle$. In this protocol, one would measure the auxiliary qubit and apply a Pauli operation that depends on the measurement outcome $m$. The measurement and conditional operation (green dashed box) can be replaced with a combination of CNOT-gates (lower left), such that no mid-circuit measurements or feed-forward operations are required. We shift this scheme to the logical level by replacing the data qubit with one logical qubit of the $[[8, 3, 2]]$-code and the auxiliary qubit with one logical qubit of a $[[4, 2, 2]]$-code, which supports a natively transversal $H_{\mathrm{L}}$-gate, up to a simple relabeling. }
	\label{fig:gates_832}
\end{figure}

Our FT logical state teleportation scheme can, in principle, be scaled to higher-distance surface codes, which is discussed further in App.~\ref{app:anticpiated_teleportation}. Here, the key idea is to use $d$ representations of logical operators on a distance-$d$ code to ensure that no weight-$d$ fault leads to a logical failure.

\subsection{FT toolbox for universal operations on the $[[8, 3, 2]]$-code}~\label{sec:832_code}

In this section, we discuss circuit constructions for a FT universal gate set on an eight-qubit error-detecting code, which we use to implement Grover's search algorithm on three logical qubits experimentally in the following section. 
The $[[8, 3, 2]]$-code is the smallest instance of a three-dimensional color code~\cite{bombin2007topological, kubica2015unfolding, campbell2016smallest, wang2024fault} that encodes $k = 3$ logical qubit in $n = 8$ physical qubits and has distance $d = 2$, meaning that any single error can be detected. The $X$-stabilizer and a $Z$-stabilizer of this code have support on all eight qubits, while three additional weight-4 $Z$-stabilizers are defined on three faces of a cube, intersecting on edges, as shown in Fig.~\ref{fig:gates_832}\textbf{a}. The three logical Pauli $X$-operators of this code have support on the weight-4 faces of the cube, while the logical $Z$-operators are defined on edges of weight 2. 

The $[[8, 3, 2]]$-code supports a transversal non-Clifford gate~\cite{campbell2016smallest, wang2024fault}: the CC$Z$-gate can be implemented by applying single-qubit $T$- and $T^{\dag}$-gates to individual qubits as illustrated in Fig.~\ref{fig:grover}\textbf{a}, such that errors do not propagate within one code block. A logical CNOT-gate between qubits, that are encoded \textit{within the same} encoded block, can be implemented by swapping pairs of qubits. In the following, we implement these CNOT-gates within one block by relabeling pairs of qubits, which does not require any physical gate operations. 
The $[[8, 3, 2]]$-code has in the past been used for multiple experimental demonstrations~\cite{bluvstein2024logical, wang2024fault, honciuc2024implementing}. 
Recent theoretical works have proposed constructions for measurement-free, FT universal quantum computing~\cite{butt2024measurement, veroni2024universal}, but require a substantial overhead in gate operations and qubit count. 
Here, we introduce an implementation of a $H_\mathrm{L}$-gate for the $[[8, 3, 2]]$-code that does not rely on mid-circuit measurements or feed-forward operations, and, together with the CC$Z$-gate, completes a FT universal gate set. 
Our construction for the FT single-qubit logical $H_\mathrm{L}$-gate is based on state injection~\cite{bravyi2005universal}. State injection makes use of a suitable resource state~\cite{goto2016minimizing}, that is injected onto the data qubit by, first, entangling the two qubits, then, measuring the resource qubit and, finally, applying a Clifford operation to the data qubit conditioned on the measurement outcome in the second step. Fig.~\ref{fig:gates_832}\textbf{b}1 shows the circuit that may be used to apply a $H_{\mathrm{L}}$-gate to a state $|\psi\rangle_\mathrm{L}$ by means of state injection. Here, an auxiliary qubit is prepared in $|+\rangle$ as a resource state, then entangled with the data qubit with a combination of a CNOT- and a C$Z$-gate. Finally, the auxiliary qubit is measured in the $X$-basis and either a Pauli $X$- or $Z$-flip is applied to the data qubit, depending on the measurement outcome $m$. 

We now replace the measurement of the auxiliary qubit with a coherent feedback operation comprising two CNOT-gates, as shown in Fig.~\ref{fig:gates_832}\textbf{b}2. A measurement can always be replaced with a quantum circuit ~\cite{Nielsen_and_Chuang, gottesman2016surviving}, but does not automatically obey fault-tolerant circuit design principles. 
In our circuit construction, we have to apply $H$-gates to the auxiliary qubit in order to achieve the desired $H$-gate injection to the data qubit. If both qubits corresponded to logical qubits of the same code, there would be no benefit in using this approach, because it would require a $H_\mathrm{L}$-gate in order to inject one. We therefore use different types of codes to inject the desired gate operation. Specifically, we consider the three encoded qubits of the $[[8, 3, 2]]$-code and inject a $H_\mathrm{L}$-gate onto one of the logical qubits by means of an auxiliary $[[4, 2, 2]]$-code prepared in $|+$$0\rangle_\mathrm{L}$ as a resource state. 
In this circuit, we require a CNOT-gate that acts on two logical qubits that are encoded in two different code blocks. This gate can be implemented with a non-transversal, yet FT, gate implementation implying that any single fault that may propagate through the full circuit remains detectable afterwards. This logical inter-block CNOT-gadget and the full circuit for the injection of a $H_\mathrm{L}$-gate are depicted in App. Fig.~\ref{fig:H_injection_circuit}. Our implementation of the FT logical $H_{\mathrm{L}}$-gate requires 4 auxiliary qubits and 26 two-qubit gates. 

We perform experimental logical state tomography for each logical qubit considering FT logical state initialization, the single-logical $H_\mathrm{L}$-gate and the transversal CC$Z_{\mathrm{L}}$-gate on the $[[8, 3, 2]]$-code. We achieve fidelities of up to 81(3)\% for $H_{\mathrm{L}}$ on logical qubit 0, accepting 10\% of the runs after postselection. Moreover, we find fidelties  between 65(6)\% and 99.89(14)\% for the two idling logical qubits, depending on the logical input state and its sensitivity to dephasing. All results are shown and further analyzed in App.~\ref{app:logical_performance_832}. Notably, we identify dephasing of idling qubits as a major error source, which we estimate to account for almost two-thirds of the overall logical error rate, as discussed further in App.~\ref{app:error_budget}. 

The presented FT universal gate set on the $[[8, 3, 2]]$-code unlocks the capability to run minimal logical algorithms without relying on explicit mid-circuit measurement or feed-forward operations. In the next step, we use it to implement a FT Grover search on three logical qubits encoded in the $[[8, 3, 2]]$-code. 

\subsection{Grover search on logical qubits}~\label{sec:grover_search}

\begin{figure*}[!tb]
	\centering
	\includegraphics[width=160mm]{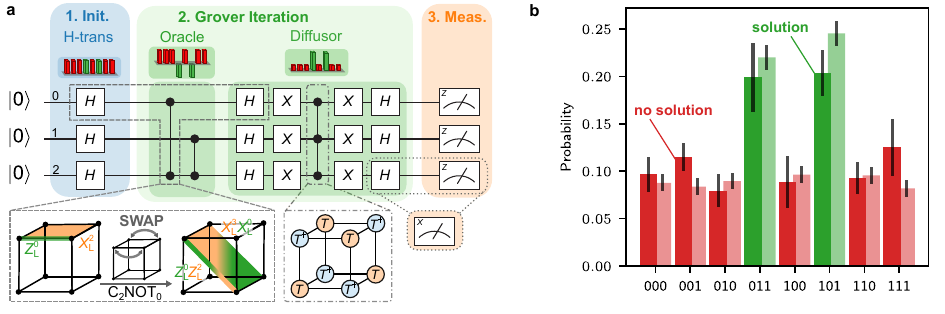}
 \caption{\justifying \textbf{Two-solution Grover search on a database of size $N = 2^3$. } \textbf{a,} We recompile the Grover search algorithm that includes a phase oracle \cite{grover1997quantum, figgatt2017complete} into the available FT gate operations within one $[[8, 3, 2]]$-code \{$H_\mathrm{L}$, CNOT$_\mathrm{L}$, CC$Z_\mathrm{L}$\}. \textbf{b,} Experimentally obtained (dark) and numerically simulated (light) probabilities for each possible solution. The two solutions 011 and 101 can be clearly distinguished, and the total probability to find one of the two solutions (green) is $p_{\mathrm{success}} = p_{011} + p_{101} = 0.40(4)$. 
 }
	\label{fig:grover}
\end{figure*}

Grover's search algorithm~\cite{grover1997quantum, boyer1998tight} enables quantum computers to search through unsorted databases significantly more efficiently than classical methods. It achieves a quadratic speedup by reducing the number of queries required to find a desired item, and can be used as a subroutine for other quantum algorithms~\cite{chakrabarty2017dynamic, chakrabarty2017dynamic, roget2020grover, tezuka2022grover, nagib2025efficient}. Grover's algorithm consists of three steps \cite{Nielsen_and_Chuang, qiskit_textbook}:
\begin{enumerate}[leftmargin=1.5em]
    \item \textit{Initialization:} Prepare all qubits in an equal-weight superposition of the computational basis states with the Hadamard transform,~i.e.~apply single-qubit $H$-gates to all qubits.
    \item \textit{Grover iteration:} Perform (a) and (b) $j$ times to amplify the amplitude of the solution-states $s$:
    \begin{enumerate}
        \item Apply an oracle operator $O$, that marks the solutions by flipping the sign of these states.
        \item Apply a diffusion operator $D$ that reflects the state about the initial state. 
    \end{enumerate}
    \item \textit{Measurement:} Measure the qubit register in the computational basis.
\end{enumerate}

We implement Grover's algorithm on three logical qubits, thus searching a database of size $N = 2^n = 8$ bits. As an example, we consider the phase oracle that marks  $s = 2$ solution-states $|011\rangle$ and $|101\rangle$. 
In this setting, the probability to find a solution after one Grover iteration in a noise-free setting is $1$, which is discussed further in App.~\ref{app:grover_search}. 
The optimal classical search corresponds to performing a single query, followed by a random guess, and the probability to find a solution in this case is $s/N + (N-s)/N \cdot s/(N-1) \approx 0.46$ in our case. Grover's search algorithm has been implemented on physical qubits on trapped ions~\cite{figgatt2017complete, brickman2005implementation}, superconducting architectures~\cite{abughanem2025characterizing, roy2018programmable, dicarlo2009demonstration}, on spin qubits in silicon~\cite{thorvaldson2025grover}, and on molecules using NMR techniques~\cite{chuang1998}. It has also been realized on two logical qubits encoded in a $[[4, 2, 2]]$-code~\cite{pokharel2024better} searching a database of $N = 4$, which does not require a universal set of gates, but can be realized with Clifford gate operations only. 

We implement the three-qubit Grover's algorithm on logical qubits encoded in the $[[8, 3, 2]]$-code by utilizing the universal FT gate set \{$H_\mathrm{L}, \mathrm{CNOT_{\mathrm{L}}}, \mathrm{CC}Z_{\mathrm{L}}\}$. We recompile the initial circuit~\cite{Nielsen_and_Chuang, grover1997quantum, figgatt2017complete} into the available FT gates introduced in the previous section, as shown in Fig.~\ref{fig:grover}\textbf{a}. We then implement this circuit on our experimental trapped-ion quantum processor, accompanied by numerical simulations according to a multi-parameter noise model specified in App.~\ref{app:anticpiated_teleportation}. Fig.~\ref{fig:grover}\textbf{b} shows the determined probabilities for each of the eight possible final states, two of which correspond to the correct solution-states as marked in green. The total probability to find a solution using the experimental data is $p_{\mathrm{success}} = p_{011} + p_{101} = 0.40(4)$. 
This overall probability to find a solution in a single shot is  slightly lower than the optimal classical probability of 0.46, as determined above. However, as discussed further in App.~\ref{app:projected_performance}, only slight enhancements to the current setup are sufficient to outperform the optimal classical algorithm. 
Numerical simulations show that reducing, e.g., the two-qubit-gate error rate by 1\% to $p_2 \approx 0.015$, which has been demonstrated on experimental trapped-ion platforms~\cite{moses2023race, ballance2016high, Chen2024benchmarkingtrapped}, leads to an overall success rate of $\approx 0.52$, which clearly outperforms the optimal classical strategy. 
Instead of reducing $p_2$, also extending the coherence time to $T_2 = $\SI{100}{\milli\second}, which has been shown in independent technical demonstrations~\cite{harty2014high, ruster2016long, wang2021single, pino2021demonstration, debnath2016demonstration}, leads to a success probability of $p_{\mathrm{success}} \approx 0.67$. This demonstrates that for only slightly smaller error rates on idling qubits and two-qubit gate operations, a regime where the measurement-free quantum algorithm outperforms its classical counterpart is reachable today.

Our scheme for Grover's algorithm can be scaled to a larger search space, provided enough qubits and sufficiently reliable gate operations are available. One can implement the FT gate set \{$H_\mathrm{L}$, CNOT$_\mathrm{L}$, CC$Z_\mathrm{L}$\} on logical qubits encoded within one $[[8, 3, 2]]$ block. In addition, one can apply an inter-block CNOT-gate between two logical qubits of two distinct $[[8, 3, 2]]$-codes~\cite{hangleiter2024fault}. These operations enable the implementation of an oracle and the amplification on more than three qubits by decomposing the required gates into the available gate sets~\cite{Nielsen_and_Chuang}.

\subsection{Outlook}

In this work, we introduce and experimentally implement a complete toolbox of operations for fault-tolerant (FT) universal quantum computing without mid-circuit measurements. Our work presents the first experimental realization of a FT universal gate set that operates without mid-circuit measurements and marks the FT implementation of Grover's algorithm on a search space of up to $N = 8$ on encoded logical qubits, demonstrating for the first time a FT logical algorithm without mid-circuit measurements. 

Our schemes are tailored towards trapped-ion architectures that provide all-to-all connectivity~\cite{reichardt2024demonstration, pino2021demonstration, moses2023race, jain2024penning}, but they can be analogously implemented on other architectures. For example, neutral atom platforms have demonstrated the capabilities required for implementing the presented code constructions~\cite{evered2023high, bluvstein2024logical%hangleiter2024fault
}. These architectures offer long-range connectivity and high-fidelity single- and two-qubit gates, while mid-circuit measurements and real-time feedback are still experimentally demanding due to relatively long measurement times~\cite{evered2023high, bluvstein2024logical, radnaev2024universal, tsai2024benchmarking}. These features make our measurement-free implementations ideal candidates for neutral atom platforms, potentially enhancing performance by avoiding costly circuit components. 

Future work will include the analysis of our protocols for higher-distance codes, as outlined above, and the investigation of thresholds and required overheads in terms of qubit count and gate operations, including extensions to fault-tolerant realizations under restricted qubit-connectivity~\cite{zen2024quantum}. 
Moreover, we have identified dephasing on idling qubits during two-qubit gates as a major logical-error source in our experimental demonstration. Further adjustment of our schemes to a biased noise setting~\cite{bonilla2021xzzx, brechtelsbauer2025measurement}, which is often given in  experimental architectures~\cite{bluvstein2024logical, pogorelov2025experimental, radnaev2024universal}, could therefore potentially boost the performance while reducing overheads. 

Our work presents the first demonstration of \textit{measurement-free} fault-tolerant quantum computation and lays the ground for further exploring the full potential of this new paradigm of fault-tolerant quantum information processing without mid-circuit measurements. 

\section*{Acknowledgments}
We gratefully acknowledge support by the European Union’s Horizon Europe research and innovation program under Grant Agreement Number 101114305 (“MILLENION-SGA1” EU Project) (C.D.M, T.M., M.Meyer, M.M\"{u}ller), the US Army Research Office through Grant Number W911NF-21-1-0007 (F.B., C.D.M, T.M., M.M\"{u}ller), the Austrian Research Promotion Agency under Contract Number 897481 (HPQC) (T.M.) supported by the European Union – NextGenerationEU, the Austrian Science Fund (FWF Grant-DOI 10.55776/F71) (SFB BeyondC) (T.M.), Intelligence Advanced Research Projects Activity (IARPA), under the Entangled Logical Qubits program through Cooperative Agreement Number W911NF-23-2-0216 (I.P., F.B., T.M., M.M\"{u}ller). 
We further receive support from the IQI GmbH, and by the German ministry of science and education (BMBF) via the VDI within the project IQuAn (M.M\"{u}ller), by the European ERC Starting Grant QNets
through Grant No. 804247 (M.M\"{u}ller) and by the Deutsche Forschungsgemeinschaft (DFG, German Research Foundation) under Germany’s Excellence Strategy ‘Cluster of Excellence Matter and Light for Quantum Computing (ML4Q) EXC 2004/1’ 390534769 (M.M\"{u}ller).

The views and conclusions contained in this document are those of the authors and should not be interpreted as representing the official policies, either expressed or implied, of IARPA, the Army Research Office, or the U.S. Government. The U.S. Government is authorized to reproduce and distribute reprints for Government purposes notwithstanding any copyright notation herein.

We acknowledge computing time provided at the NHR Center NHR4CES at RWTH Aachen University (Project No. p0020074) (F.B., M.M\"{u}ller). This is funded by the Federal Ministry of Education and Research and the state governments participating on the basis of the resolutions of the GWK for national high-performance computing at universities. 

\section*{Data availability}
The data provided in the figures in this article, the explicit circuits and the code that was used to simulate the presented protocols are available at \url{https://doi.org/10.5281/zenodo.15747203}. 

\section*{Author contributions:}
F.B. developed the presented protocols and performed the numerical simulations. I.P. and R.F. implemented the presented protocols on the experimental setup and performed experiments. I.P., R. F., A. S. and M. Meyer built and maintained the experimental setup. F.B., I.P. and R.F. analyzed results. F. B. and I. P. wrote the manuscript, with contributions from all authors. T. M. and M.M\"{u}ller supervised the project.

\section*{Competing interests:}
T.M. is connected to Alpine Quantum Technologies GmbH, a commercially oriented quantum computing company. The remaining authors declare no competing interests.

\clearpage

\appendix

\section{Qubit reset} \label{app:reset}

\begin{figure*}[!tb]
    \centering
    \includegraphics[width=160mm]{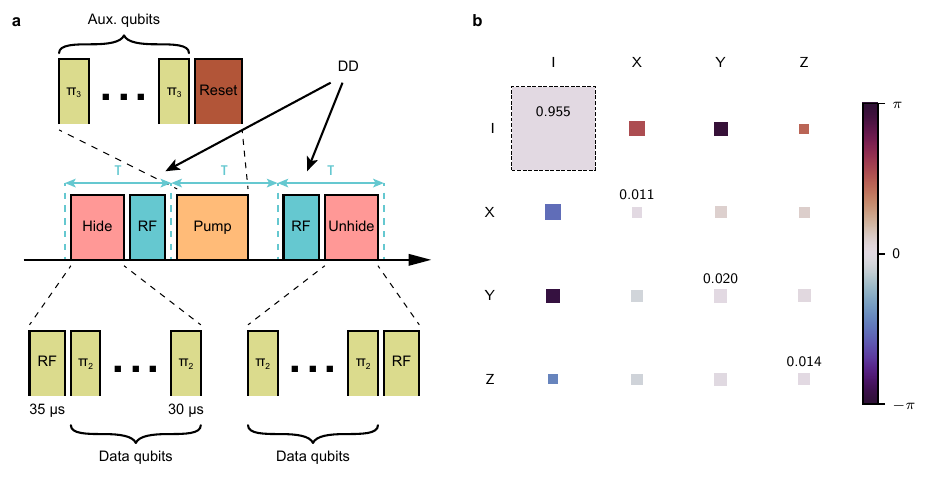}
    \caption{\justifying \textbf{Experimental details of the qubit reset procedure. } \textbf{a,} The pulse sequence implementing the qubit reset procedure. Firstly, the data qubits are hidden in the $4^2\textrm{S}_{1/2}$ manifold by applying a global RF pulse and individually addressing each data qubit with a $\pi$-pulse. After that, the auxiliary qubits are reinitialized as $\ket{0}$. Finally, the data qubits' encoding is restored to the original. The dynamical decoupling (DD) pulses are inserted to mitigate decoherence of the data qubits in the $4^2\textrm{S}_{1/2}$ manifold. \textbf{b,} The $\chi$-matrix representation of the reset procedure as the process acting on data qubits, averaged over all data qubits. The area and the color coding of the squares correspond to the absolute value and the phase of an element of the $\chi$-matrix, respectively. The dashed square represents an ideal outcome, specifically the identity process. }
    \label{fig:reset}
\end{figure*}

The qubit reset procedure allows for a selective reinitialization of some of the qubits to state $\ket{0}$. 
Physically, the reset is performed by quenching the lifetime of the $3^2\textrm{D}_{5/2}$ manifold by illuminating the ion chain with the \SI{854}{\nano\meter} laser with subsequent optical pumping to reinitialize the qubits in the $\ket{0}$ state. The data qubits are hidden in the $4^2\textrm{S}_{1/2}$ manifold $\ket{4^2\textrm{S}_{1/2}, m_J = -1/2}, \ket{4^2\textrm{S}_{1/2}, m_J = +1/2}$ by means of the electron shelving technique \cite{schindler2013quantum} during the life time quenching to preserve their state. This 
re-encoding of the data qubits in the $\ket{4^2\textrm{S}_{1/2}, m_J = -1/2}, \ket{4^2\textrm{S}_{1/2}, m_J = +1/2}$ levels instead of the $\ket{4^2\textrm{S}_{1/2}, m_J = -1/2}, \ket{3^2\textrm{D}_{5/2}, m_J = -1/2}$ levels results in a higher sensitivity to magnetic field noise and, consequently, lower coherence time. Therefore, we perform two dynamical decoupling pulses (DD) with the radio-frequency (RF) antenna driving the transition between the $\ket{4^2\textrm{S}_{1/2}, m_J = -1/2}$ and $\ket{4^2\textrm{S}_{1/2}, m_J = +1/2}$ levels. The sketch of the procedure is shown in Fig.~\ref{fig:reset}\textbf{a} while additional details can be found in \cite{postler2023demonstration}.

The reset procedure does not require recooling of the ion chain, unlike the full mid-circuit measurement, since the reset ions emit only a few photons during the procedure. Consequently, the reset is faster than our current implementation of the mid-circuit measurement (\SI{1.7}{\milli\second} vs. $\approx\SI{30}{\milli\second}$) and the preservation of the data qubit's state is higher (process fidelity 0.955(9) vs. 0.908(12)). The $\chi$-matrix for the reset procedure obtained via quantum process tomography is depicted in Fig.~\ref{fig:reset}\textbf{b}.
We make use of the full 16-ion register and use fresh auxiliary qubits as long as possible. The reset procedure is only used in our implementation of Grover's algorithm (see Fig.~\ref{fig:grover}) to reset one auxiliary qubit that is used for the FT preparation of $\ket{+00}_\mathrm{L}$ for the $[[8,3,2]]$-code (see Fig.~\ref{fig:initialization_circuit}\textbf{e}). This auxiliary qubit is later used for the mapping of one $Z$-stabilizer of the $[[8, 3, 2]]$-code, as discussed in App.~\ref{app:tomography}.

\section{Anticipated performance of measurement-free state teleportation}\label{app:anticpiated_teleportation}

\begin{figure*}[!tb]
	\centering
	\includegraphics[width=172mm]{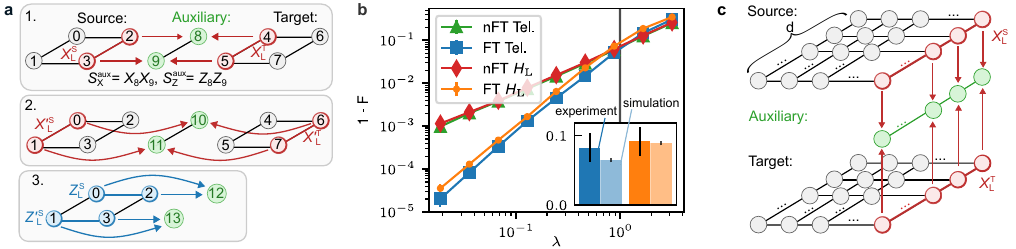}
    \caption{\justifying \textbf{Fault tolerance and scaling to higher code distances for measurement-free logical state teleportation. }\textbf{a}, We prepare an auxiliary two-qubit GHZ-state, to prevent single faults on auxiliary qubits from causing a logical failure. In addition, we map two stabilizer-equivalent representations of the joint logical operator with fully disjoint qubit support onto the auxiliary registers (1., 2.). 
    The same strategy is used in step 3, where two equivalent but fully disjoint logical $Z$-operators of the source register are mapped onto two physical qubits. \textbf{b}, Numerically determined scaling of the logical infidelity for FT and non-FT logical teleportation and the $H_\mathrm{L}$-gate, averaged over initial states $|0\rangle_L$ and $|+\rangle_L$. We fix the error parameters $\vec{p} = (p_\mathrm{1}, p_\mathrm{2}, p_\mathrm{m}, p_\mathrm{i}, p_{\mathrm{idle}})$ to experimental error rates~\cite{pogorelov2025experimental, postler2023demonstration} and scale these with a common improvement factor $\lambda$. We identify a quadratic scaling of the infidelity for the FT protocols with $\lambda$, which indicates--as expected--that no single fault leads to a logical failure. The FT teleportation protocol outperforms its non-FT counterpart already for the current experimental noise parameters ($\lambda = 1$). The inset shows the logical infidelities at $\lambda = 1$ obtained from the experiment (darker color) and numerical simulations (lighter color). \textbf{c}, Scaling measurement-free state teleportation to surface codes with higher distances $d>2$. Each string of qubits connecting opposing boundaries supports a representation of a logical $X_\mathrm{L}^\mathrm{S}$ (upper lattice) and $X_\mathrm{L}^\mathrm{T}$ (lower lattice); one exemplary representation is shown in red. There are $d$ equivalent representations that have fully disjoint support. Each one can be mapped onto an auxiliary $d$-qubit GHZ-state, and coherent feedback steps can be applied, which are controlled by the state on $d$ physical auxiliary qubits. }
	\label{fig:anticpiated_performance_tel}
\end{figure*}

The measurement-free logical teleportation schemes are made FT as illustrated in Fig.~\ref{fig:anticpiated_performance_tel}\textbf{a}. First, we prepare auxiliary two-qubit GHZ-states $|\psi_{\mathrm{aux}}\rangle = (|00\rangle + |11\rangle)/\sqrt{2}$ stabilized by $S_X^{\mathrm{aux}} = X_8 X_9$ and $S_Z^{\mathrm{aux}} = Z_8 Z_9$, which ensures that no single fault on an auxiliary qubit propagates to a logical error when the two registers are coupled. 
In addition, we map two representations of the joint logical operator $X_\mathrm{L}^\mathrm{S}X_\mathrm{L}^\mathrm{T}$, that have fully disjoint support, onto the auxiliary register such that no single fault on a data qubit leads to a logical error on the output state (panels 1. and 2. in Fig.~\ref{fig:anticpiated_performance_tel}\textbf{a}). Here, the information about each representation of the joint logical operator is stored in one physical auxiliary qubit, which then acts as a control qubit in the coherent feedback operation consisting of C$Z$-gates. The same strategy is used in step 3, where two representations of $Z_\mathrm{L}^{\mathrm{S}}$ with disjoint support are mapped onto two physical qubits. The explicit circuits can be found in App.~\ref{app:circuits}. The non-FT (nFT) counterparts of these protocols make use of a bare physical auxiliary qubit and only map a single representation of the respective operators onto this auxiliary qubit, which is then used to control the coherent feedback operation. 

In Fig.~\ref{fig:anticpiated_performance_tel}\textbf{b}, we simulate the scaling of the logical infidelity for FT and non-FT measurement-free logical state teleportation and the application of the $H_\mathrm{L}$-gate operation by means of teleportation. Here, we consider a multi-parameter noise model, attributing different error rates to each type of component in the circuits. Specifically, we consider depolarizing noise on single-qubit gates with a probability $p_1 = 3.6\cdot 10^{-3}$, two-qubit depolarizing noise on two-qubit gates with a probability $p_2 = 2.5\cdot 10^{-2}$, flipped physical qubit initializations with a probability $p_i = 3 \cdot 10^{-3}$, and flips before the final projective measurements with a probability $p_m = 3\cdot 10^{-3}$. The values of the error rates correspond to the ones in our experimental setup \cite{pogorelov2021compact, postler2022demonstration, postler2023demonstration, pogorelov2025experimental}. We implement dephasing on all idling qubits, where a $Z$-fault is applied to each idling physical qubit with a probability $p_{\mathrm{idle}} = (1 - e^{-t_{\mathrm{gate}}/T_2})$ given the gate time of the respective operation and the coherence time $T_2 = \SI{50}{\milli\second}$. Note that gates in the experimental setup can only be executed sequentially, which increases the total dephasing time. The noise channels and numerical methods are explicitly given in App.~\ref{app:numerical_methods}. We scale the error parameters $\vec{p}(\lambda) = \lambda \cdot (p_1, p_2, p_i, p_m, p_{\mathrm{idle}})$ with a common factor $\lambda$, such that $\lambda = 1$ corresponds to the set of parameters as given in the current experimental setup. As expected, the FT protocols scale quadratically with $\lambda$, indicating that the required fault-tolerance properties are fulfilled. 

Our approach for logical state teleportation without mid-circuit measurements is, in principle, scalable to higher distance $d>2$ surface codes, as illustrated in Fig.~\ref{fig:anticpiated_performance_tel}\textbf{c}. These codes have $d$ equivalent representations of the logical Pauli-operators which do not share support. For the FT mapping of the weight-$d$ logical operators, we then have to prepare the auxiliary register fault-tolerantly in a $d$-qubit GHZ-state, and apply coherent feedback steps controlled on the state of $d$ physical auxiliary qubits. 

\section{Numerical methods}\label{app:numerical_methods}
We use Monte Carlo simulations to estimate the logical infidelities of our protocols~\cite{pecos_git}. Each circuit component is modeled by first applying the respective ideal operation, followed by an error $E$ occurring with probability $p$. We simulate a depolarizing noise channel after every single- and two-qubit gate. With probabilities $p_1$ and $p_2$, an error from the respective sets is applied. These probabilities define the corresponding error channels 
\begin{align}
    \mathcal{E}_1(\rho) &= (1 - p_1)\rho + \frac{p_1}{3} \sum_{i= 1}^3 E^{i}_1 \rho E^{i}_1 \\
    \mathcal{E}_2(\rho) &= (1 - p_2)\rho + \frac{p_2}{15} \sum_{i= 1}^{15}   E_2^{i} \, \rho\, E_2^{i}.  \nonumber\label{eq:depol_single_qubit}
\end{align}
with $E_1^k \in \{X$, $Y$, $Z\}$ for $k = 1, 2, 3$ and 
$E_2^k$ $\in$ $\{IX$, $XI$, $XX$, $IY$, $YI$, $YY$, $IZ$, $ZI$, $ZZ$, $XY$, $YX$, $XZ$, $ZX$, $YZ$, $ZY\}$ for $k = 1, ..., 15$. All qubits are initialized and measured in the $Z$-basis, at the very end of the respective protocols. To simulate faults in these operations, we apply $X$-flips after initialization and before measurement, each occurring with probabilities $p_{\mathrm{init}}$ and $p_{\mathrm{meas}}$, respectively. Moreover, qubits that remain idle during gate operations may experience dephasing, which we model with the error channel
\begin{align}
    \mathcal{E}_{\mathrm{idle}}(\rho) &= (1 - p_{\mathrm{idle}})\rho + p_{\mathrm{idle}} Z\rho Z. 
\end{align}
The probability $p_{\mathrm{idle}}$ depends on the execution time $t$ of the performed gate and the qubit coherence time $T_2 = 50\,$ms 
\begin{align}
    p_{\mathrm{idle}} = \frac{1}{2} \left[1 - \mathrm{exp}\left(-\frac{t}{T_2} \right) \right]. 
\end{align}
In our simulations, we use $t_1 = $ \SI{70}{\micro\second} as the gate time of single-qubit gates and $t_2 = $ \SI{350}{\micro\second} as the gate time of two-qubit gates, as summarized in Tab.~\ref{tab:error_rates_simulation}. 

We measure the final state in the logical $X$-, $Y$-, and $Z$-basis for each protocol, as described in App.~\ref{app:tomography}, and calculate the state fidelity between the ideal logical state $\rho_1$ and the reconstructed density operator $\rho_2$ obtained after postselecting, as
\begin{align}
    F(\rho_1, \rho_2) = \mathrm{Tr}{\left[\sqrt{\sqrt{\rho_1} \rho_2 \sqrt{\rho_1} } \right]^2}.
\end{align}
We use Qiskit's Quantum Information package to calculate fidelities~\cite{Qiskit_quantuminfo}.

\begin{table}
    \centering
    \renewcommand*{\arraystretch}{1.1}
    \begin{tabular}{c|c|c}
         Operation & Error rate & Duration\\
         \hline
         Two-qubit gate & $p_{2q} = 0.025$ & \SI{350}{\micro s}\\
         \hline
         Single-qubit gate & $p_{1q} = 0.0036$ & \SI{70}{\micro s}\\
         \hline
         Measurement & $p_{\mathrm{meas}} = 0.003$& -\\
         \hline
         Preparation & $p_{\mathrm{init}} = 0.003$& -\\
    \end{tabular}
    \caption{\justifying \textbf{Error rates and duration of operations on a trapped-ion quantum processor.} These values correspond to the trapped-ion setup that was used in the experiments and are used in the numerical simulations. Furthermore, the coherence time is determined to be $T_2 =\,$\SI{50}{\milli\second} in our experimental setup. }
    \label{tab:error_rates_simulation}
\end{table}

\section{Circuits}\label{app:circuits}

\begin{figure*}[!tb]
	\centering
	\includegraphics[width=180mm]{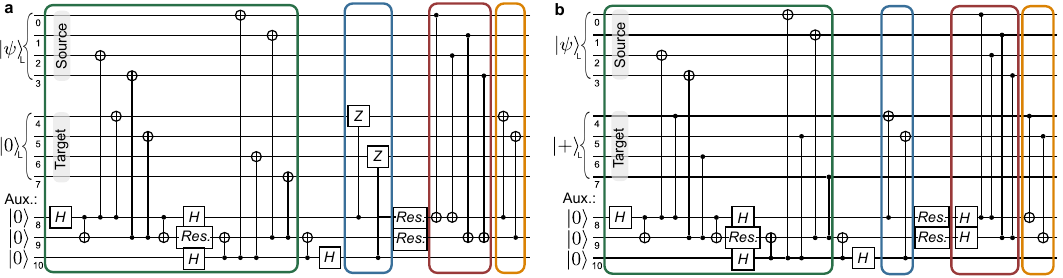}
 \caption{\justifying \textbf{Circuits for modular measurement-free logical state teleportation from a source to the target register and the application of a $H_\mathrm{L}$-gate by means of teleportation. }\textbf{a,} We first map two stabilizer-equivalent joint logical operators $X_{\mathrm{L}}^{\mathrm{S}} X_{\mathrm{L}}^{\mathrm{T}} = X_2 X_3 X_4 X_5$ and $X_{\mathrm{L}}^{\mathrm{S}} X_\mathrm{L}^\mathrm{T} = X_0 X_1 X_6 X_7$ that have fully disjoint support onto auxiliary two-qubit GHZ-states (green). Then, two C$Z$-gates are applied implementing the desired $Z_{\mathrm{L}}^{\mathrm{T}}$, if both values of the joint logical operators are in 1 (blue). Here, \textit{Res.} corresponds to a reset operation as described in App.~\ref{app:reset}, which can either be carried out explicitly by physically resetting the auxiliary qubits to the $|0\rangle$ state and reusing them afterwards, or implemented by replacing them with fresh qubits. In the second step, we map two equivalent operators $Z_{\mathrm{L}}^{\mathrm{S}} = Z_0 Z_2$ and $Z_{\mathrm{L}}^{\mathrm{S}} = Z_1 Z_3$ onto physical auxiliary qubits (red) and implement two CNOT-gates to apply $X_{\mathrm{L}}^{\mathrm{T}}$, if both auxiliary qubits are in the $|1\rangle$-state (orange). \textbf{b,} Analog circuit for applying a $H_{\mathrm{L}}$-gate to the $[[4, 1, 2]]$-code. We now prepare the target register in $|+\rangle_\mathrm{L}$, instead of $|0\rangle_\mathrm{L}$. Then we map out two disjoint, but equivalent operators $X_{\mathrm{L}}^{\mathrm{S}} Z_{\mathrm{L}}^{\mathrm{T}}$ (green), implement a coherent quantum feedback with CNOT-gates (blue), map two $Z_{\mathrm{L}}^{\mathrm{S}}$ to physical auxiliary qubits (red) and apply a coherent feedback operation with two CNOT-gates (orange). Note that the last two steps are recompiled into C$Z$- and CNOT-gates and some $H$-gates acting on the auxiliary qubits at the end are omitted, as the auxiliary qubits are disentangled from the data-qubit registers and discarded afterwards. }
	\label{fig:teleportation_circuit}
\end{figure*}

\begin{figure*}[!tb]
	\centering
	\includegraphics[width=115mm]{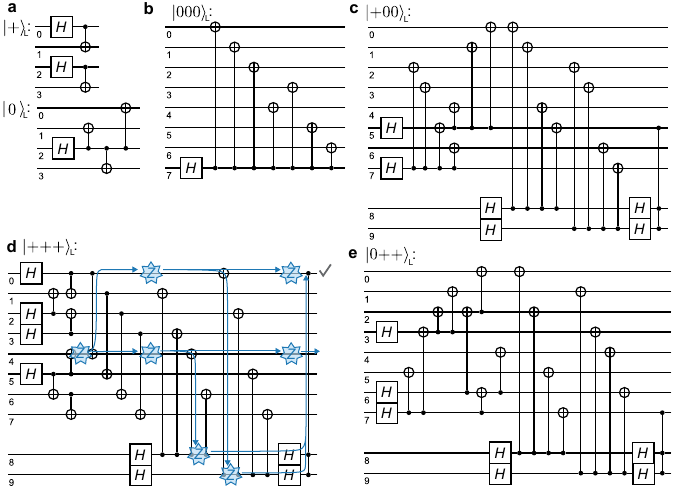}
 \caption{\justifying \textbf{Circuits for fault-tolerant logical state preparation on the $[[4, 1, 2]]$ and $[[8, 3, 2]]$ code.} \textbf{a,} FT initialization circuit for $|+\rangle_{\mathrm{L}}$ (top) and $|0\rangle_{\mathrm{L}}$ (bottom) for the $[[4, 1, 2]]$-code. The first circuit is also used to initialize $|+$$0\rangle_{\mathrm{L}}$ on the $[[4, 2, 2]]$-code. \textbf{b-e,} FT initialization circuits for $|000\rangle_{\mathrm{L}}$, $|+$$00\rangle_{\mathrm{L}}$, $|+$$++\rangle_{\mathrm{L}}$ and $|0+$$+\rangle_{\mathrm{L}}$ for the $[[8, 3, 2]]$-code. For the last three, we add a coherent correction step in the end: two equivalent logical $X_{\mathrm{L}}$-operators that have fully disjoint support are mapped onto two physical auxiliary qubits, and a CC$Z$-operation is applied to correct a potentially dangerous weight-2 error resulting from a single fault into a detectable weight-1 error. For example, a single $Z$-fault in the circuit for the $|+$$++\rangle_{\mathrm{L}}$ initialization (\textbf{d}) on qubit 4 (marked in blue) may propagate onto qubits 0 and 4, which directly corresponds to a logical error $Z_L^1$. In the measurement-free verification step, this error configuration propagates further onto \textit{both} auxiliary qubits. The CC$Z$-gate in the end applies $Z_0$ to the data qubits, as both auxiliary qubits have been flipped, effectively removing one of the errors and leaving the detectable error configuration $Z_4$. }
	\label{fig:initialization_circuit}
\end{figure*}

\begin{figure*}[!tb]
	\centering
	\includegraphics[width=115mm]{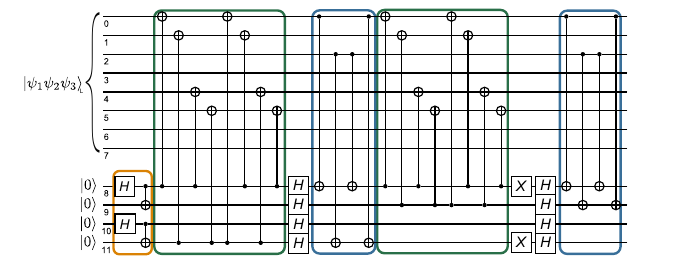}
 \caption{\justifying \textbf{Circuit for a single-logical qubit $H_\mathrm{L}$-gate on the $[[8, 3, 2]]$ code, as illustrated in Fig.~\ref{fig:gates_832}b3.} We first prepare the auxiliary register in the $|+$$0\rangle_{\mathrm{L}}$-state of the $[[4, 2, 2]]$-code (orange). Then, we apply the inter-block CNOT-gate (green) where the control-bit corresponds to the first qubit of the $[[4, 2, 2]]$-block, and the target qubit to the first qubit of the $[[8, 3, 2]]$-code. The inter-block CNOT-gate is not transversal but FT in the sense that any single fault may propagate, but is still detectable in the end. In the next step, the $H^{\otimes 2}_{\mathrm{L}}$-gate is applied to the auxiliary register. This step includes an additional SWAP operation, which is absorbed by the following gates, as the physical qubits are simply relabeled accordingly. 
 We then apply an inverted CNOT-gate (blue), that is controlled by the first logical qubit of the $[[8, 3, 2]]$ and acts on the first one of the $[[4, 2, 2]]$-code. The two inter-block logical CNOT-gates are not symmetric, but the implementation depends on the orientation of the gate. After again applying C$_{\mathrm{aux, 1}}$NOT$_{\psi 1}$ (green), and $H^{\otimes 2}_{\mathrm{L}} X_{\mathrm{L}}^1 $ to the logical auxiliary qubits, we finally apply the last logical gate C$_{\psi 1}$NOT$_{\mathrm{aux 1}}$ (blue). }
	\label{fig:H_injection_circuit}
\end{figure*}

Figure~\ref{fig:teleportation_circuit} shows the explicit circuits that were implemented for FT logical state teleportation discussed in Sec.~\ref{sec:MF_teleportation}. Figure~\ref{fig:initialization_circuit} shows the circuit constructions for the FT logical state initializations on the $[[4, 1, 2]]$- and the $[[8, 3, 2]]$-code without mid-circuit measurements as implemented in the demonstrated protocols. In these circuits, fault tolerance is maintained even without measurements by means of a flag-qubit-controlled reduction of potentially dangerous weight-2 errors to weight-1 configurations. Figure~\ref{fig:H_injection_circuit} depicts the circuit construction for the application of a single-logical $H_\mathrm{L}$-gate on the $[[8, 3, 2]]$-code. 

% teleportation and H with teleportation

% initializations

% H-injection

\begin{figure}[!tb]
    \centering
    \includegraphics[width=90mm]{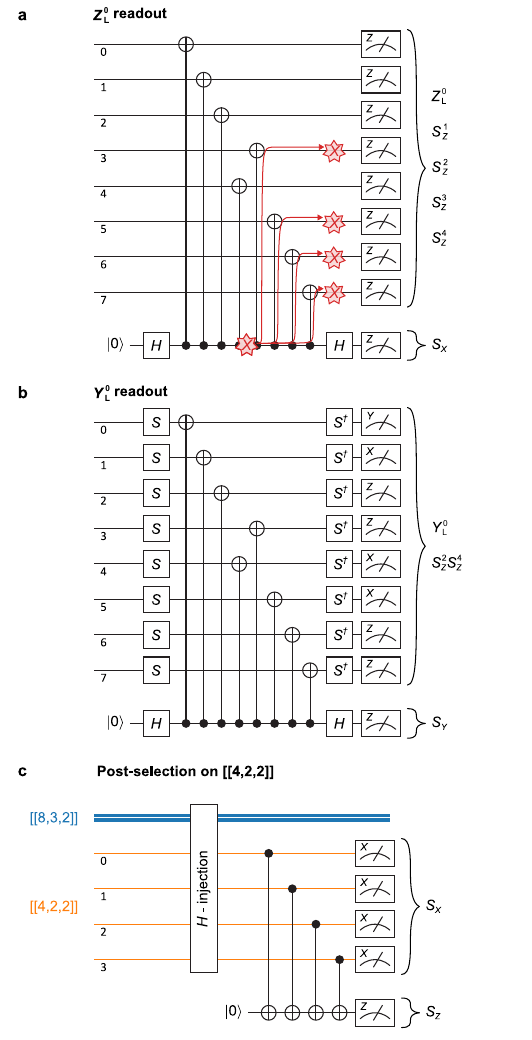}
    \caption{\justifying \textbf{FT measurement instructions for the $[[8,3,2]]$-code.} \textbf{a,} Circuit for the FT mapping of the $S_X$-stabilizer onto a physical auxiliary qubit, required for FT measurements in the $Z$-basis. A single $X$-fault on the auxiliary qubit may propagate to a weight-4 error configuration, as for example to $X_3 X_5 X_6 X_7$. This error is detected by~e.g. the $Z$-stabilizer $S_Z^1 = Z_0 Z_1 Z_2 Z_3$. \textbf{b,} Exemplary circuit for FT measurement in the $Y$-basis. We first map the $Y$-stabilizer $S_Y$ onto a physical auxiliary qubit and subsequently measure the physical data qubits in different bases, as indicated, to infer $Y_{\mathrm{L}}^0$. \textbf{c,} Circuit for stabilizer extraction on the logical auxiliary $[[4, 2, 2]]$-code. We map the $Z$-stabilizer onto an auxiliary qubit and then destructively measure the physical qubits in the $X$-basis.}
    \label{fig:832_code_readout}
\end{figure}

\section{Tomography}\label{app:tomography}

\begin{figure}[!tb]
    \centering
    \includegraphics[width=90mm]{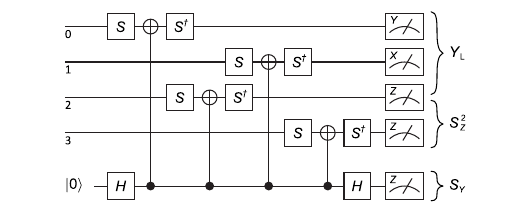}
    \caption{\justifying \textbf{FT measurement of $Y_{\mathrm{L}}$.} We first map the $Y$-stabilizer onto an auxiliary qubit and then measure the physical qubits in the $Y$-, $X$- and $Z$-basis to fault-tolerantly measure $Y_{\mathrm{L}}$. }
    \label{fig:Y_stab_mapping_412_code}
\end{figure}

\noindent \textbf{$[[4, 1, 2]]$-code}\\
We perform logical state tomography for two logical input states $|0\rangle_{\mathrm{L}}$ and $|+\rangle_{\mathrm{L}}$ considering logical state preparation, state teleportation and the application of a $H_\mathrm{L}$-gate on the $[[4, 1, 2]]$-code, as shown in Fig.~\ref{fig:teleportation}. To this end, we measure in the $X$-, $Y$-, and $Z$-basis to extract the respective logical expectation values. For measurements in the $X$-basis, we measure all physical qubits in the $X$-basis in the end and infer the logical $X$-operator and the $X$-type stabilizer from this measurement. Analogously, we can extract the logical $Z$-operator and the $Z$-type stabilizers for measurements of all physical qubits in the $Z$-basis. However, we cannot simply determine the required stabilizers and the logical value at the same time for measurements in the $Y$-basis, because they share support but are of different Pauli-type, as for example $Y_{\mathrm{L}} = Y_0 X_1 Z_2$ and the $Y$-type stabilizer $S_Y = Y_0 Y_1 Y_2 Y_3$. We therefore map out the $Y$-stabilizer onto a physical auxiliary qubit with the circuit shown in Fig.~\ref{fig:Y_stab_mapping_412_code}. Here, the gate ordering ensures that no hook error, i.e. a fault on an auxiliary qubit that may propagate onto multiple data qubits, leads to a logical flip in the subsequent measurement, as any single propagated fault is still detected in the end by a $Z$-stabilizer. We finally measure the 4 qubits in the $Y$-, $X$-, and $Z$-basis, allowing us to determine $Y_{\mathrm{L}}$ and one additional $Z$-stabilizer $S_Z^2 = Z_2 Z_3$. \\

\noindent \textbf{$[[8, 3, 2]]$-code}\\
We perform logical state tomography for each state prepared with the specified protocol,~i.e. logical state preparation and the logical operations $H_{\mathrm{L}}$ and CC$Z_{\mathrm{L}}$ on $[[8, 3, 2]]$ for different input states, as shown in Fig.~\ref{fig:logical_operations_results}. We consider each individual logical qubit and perform tomography for each one independently. 
For measurements in the $X$-basis, we measure all physical qubits in the $X$-basis and determine the three logical Pauli-operators $X_{\mathrm{L}}^0$, $X_{\mathrm{L}}^1$, $X_{\mathrm{L}}^2$ and the $X$-stabilizer $S_X$, as defined in Fig.~\ref{fig:gates_832}\textbf{a}. 
Analogously, we extract the logical $Z$-operators simultaneously, along with the $Z$-stabilizers, when measuring all physical qubits in the $Z$-basis. In this case, we additionally map out the $X$-stabilizer $S_X$ onto an auxiliary qubit when performing the $H_{\mathrm{L}}$-gate to achieve fault tolerance. The circuit that is used for measurements in the $Z$-basis is shown in Fig.~\ref{fig:832_code_readout}\textbf{a}. In this circuit, a single fault may propagate as illustrated in red, but is detected by the $Z$-stabilizers afterwards. 

For measurements in the $Y$-basis, we have to take into account that the different logical operators may share support but are of different Pauli-type, as for example $Y_{\mathrm{L}}^0 = Y_0 X_1 Z_2 X_4 X_5$ and $Y_{\mathrm{L}}^1 = Y_0 X_1 X_2 X_3 Z_4$, so they cannot be extracted simultaneously in a single measurement. We therefore perform three sets of independent experiments and determine $Y_{\mathrm{L}}^0$, $Y_{\mathrm{L}}^1$ and $Y_{\mathrm{L}}^2$ individually. For measurements in the $Y$-basis, we also map the $Y$-stabilizer $S_Y = Y_0 Y_1 Y_2 Y_4 Y_3 Y_5 Y_6 Y_7$ onto an auxiliary qubit when performing the $H_{\mathrm{L}}$-gate. We then measure the physical qubits in different bases to extract the respective logical $Y$-operator and one additional $Z$-stabilizer. An exemplary circuit that is used for measurement in the $Y$-basis for the extraction of $Y_{\mathrm{L}}^0$ is shown in Fig.~\ref{fig:832_code_readout}\textbf{b}. When extracting $Y_{\mathrm{L}}^1$ and $Y_{\mathrm{L}}^2$, we measure the physical qubits in the bases $Y_0  X_1 X_2 X_3 Z_4 Z_5 Z_6 Z_7$ and $Y_0 Z_1 X_2 Z_3 X_4 Z_5 X_6 Z_7   $, respectively. 

Moreover, the logical auxiliary qubit is still intact after performing the single-logical $H_{\mathrm{L}}$-gate. We also projectively measure the logical auxiliary qubit, extract the stabilizers of the $[[4, 2, 2]]$-instance, and postselect for a trivial syndrome to increase the fidelities in our protocols. Here, we map the $S_Z$-stabilizer of the logical auxiliary qubit onto another physical auxiliary with the circuit shown in Fig.~\ref{fig:832_code_readout}\textbf{c}. 

When we run the full logical Grover search algorithm on the three qubits of the $[[8, 3, 2]]$-code, we additionally map two $Z$-stabilizers $S_Z^1 = Z_0 Z_1 Z_2 Z_3$ and $S_Z^2S_Z^3 = Z_1 Z_2 Z_5 Z_6$ onto physical auxiliary qubits in the end in order to maintain fault tolerance.

\section{Number of measurements}\label{app:number_measurements}

In the tomography experiments described in App.~\ref{app:tomography}, each logical state was measured in three measurement bases $\{X, Y, Z\}$ with the same number of measurements for each basis. The teleportation experiment with the $[[4,1,2]]$-code (see Fig.~\ref{fig:teleportation}) took 40000 shots for each logical state per measurement basis. The initialization and logical operations with the $[[8,3,2]]$-code (see Fig.~\ref{fig:logical_operations_results}) took 7500 shots for each logical state and logical qubit per measurement basis. The Grover's algorithm demonstration (see Fig.~\ref{fig:grover}) took 37500 shots per measurement basis. All data sets were split into 12 equal subsets, the tomography was performed for every subset, yielding 12 values for the fidelity for every experiment. The  final fidelity numbers are the mean and the standard deviation of these 12 values.
\clearpage

\section{Performance of the FT universal gate set on the $[[8, 3, 2]]$-code}\label{app:logical_performance_832}

\begin{figure*}[!tb]
	\centering
	\includegraphics[width=180mm]{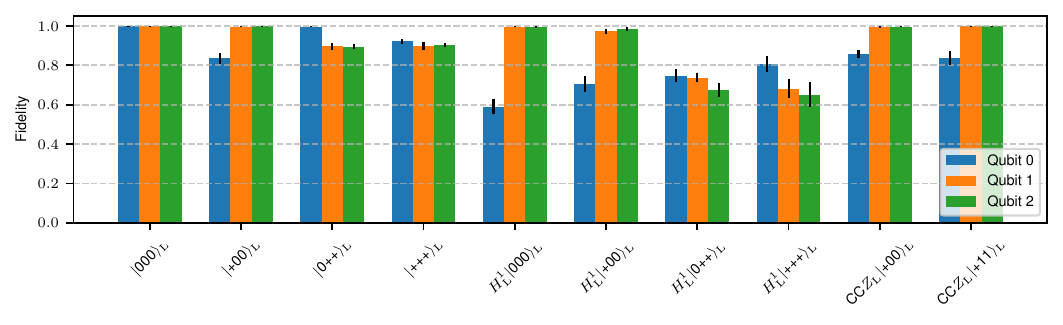}
 \caption{\justifying \textbf{Fidelities for different logical input states and non-trivial logical operations.} Experimentally obtained logical state fidelities for state initialization, a single-logical $H_\mathrm{L}$-gate and the transversal CC$Z_\mathrm{L}$-gate for the $[[8,3,2]]$ code. }
	\label{fig:logical_operations_results}
\end{figure*}

\begin{figure*}[!tb]
    \centering
    \includegraphics[width=180mm]{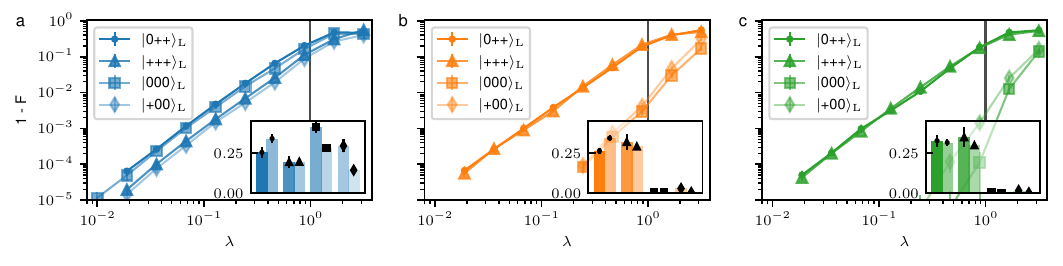}
    \caption{\justifying \textbf{Scaling of the logical infidelity for different logical qubits during the single logical $H_\mathrm{L}$-gate. } We fix the error parameters $\vec{p} = (p_\mathrm{1}, p_\mathrm{2}, p_\mathrm{m}, p_\mathrm{i}, p_{\mathrm{idle}})$ and scale these with $\lambda$. $\lambda = 1$ corresponds to the current values in the experimental setup. We determine the scaling of the logical infidelity for the logical qubit 0(\textbf{a}), on which the $H_\mathrm{L}$ is applied, and idling logical qubits 1(\textbf{b}) and 2(\textbf{c}). The inset shows the logical infidelities at $\lambda = 1$ obtained from the experiment (darker color) and from numerical simulations (lighter color). }
    \label{fig:scaling_H_injection}
\end{figure*}

Fig.~\ref{fig:logical_operations_results} shows the logical state fidelities that were obtained experimentally for FT logical state initialization, the single-logical $H_\mathrm{L}$-gate and the transversal CC$Z_{\mathrm{L}}$-gate on the $[[8, 3, 2]]$-code. 
We find that fidelities are higher if the final target state is a $Z$-eigenstate, as opposed to an $X$-eigenstate, due to dephasing, which does not affect the fidelity for $Z$-eigenstates. Additionally, postselection based on the four $Z$-stabilizers is more selective than only a single $X$-stabilizer, which boosts the fidelities in these cases. The degree of postselection is reflected in the acceptance rates: the average acceptance rates in the experiment[simulation] after the state initialization are 0.6[0.64] in the $X$-basis, 0.48[0.53] in the $Y$-basis, and 0.3[0.48] for measurements in the $Z$-basis; the numbers in brackets indicate the acceptance rate obtained in the simulation. After the injection of a $H_\mathrm{L}$-gate, these are 0.3[0.2] for measurements in the $X$-basis, 0.2[0.13] in the $Y$-basis and 0.1[0.07] in the $Z$-basis.
The fidelities for the state initialization of $|+$$00\rangle_{\mathrm{L}}$ and CC$Z_\mathrm{L}|+$$00\rangle_{\mathrm{L}}$ agree with each other within the given uncertainty interval, since the CC$Z_{\mathrm{L}}$ consists entirely of virtual $Z$-rotations, thus no additional operations are physically applied to the qubits. 
\onecolumngrid
\twocolumngrid
Fig.~\ref{fig:scaling_H_injection} shows the simulated scaling of the logical infidelity for the logical $H_{\mathrm{L}}$-gate on the $[[8, 3, 2]]$-code for each logical qubit. We scale the noise parameters $\vec{p}(\lambda) = \lambda \cdot (p_1, p_2, p_i, p_m, p_{\mathrm{idle}})$ given the same values as specified in App.~\ref{app:anticpiated_teleportation}, such that $\lambda = 1$ corresponds to the set of parameters as given for the current experimental setup. The inset shows the state fidelities for the different logical input states obtained from experiment and simulation. 

We find that the fidelities of the first logical qubit obtained from simulation, shown in blue in Fig.~\ref{fig:scaling_H_injection}\textbf{a}, differ from the experimental result by more than 14\% for logical states $|+00\rangle_{\mathrm{L}}$ and  $|000\rangle_{\mathrm{L}}$. We attribute this deviation to \textit{global} dephasing effects due to random fluctuations in the effective magnetic field that act on all physical qubits simultaneously~\cite{pal2022relaxation}, instead of locally and uncorrelated on each individual qubit. The effect of this global dephasing on the eight-qubit state can be estimated by considering the explicit basis states, for example
\begin{align}
    |000\rangle_{\mathrm{L}} = \frac{1}{\sqrt{2}} (|00000000\rangle + |11111111\rangle). 
\end{align}
Local dephasing on this state leads to decay of the off-diagonal elements of the density matrix with a factor of $e^{-\Delta n /2 \cdot \gamma t}$, where $\gamma$ is a decay constant and $t$ is time. $\Delta n$ is the number of positions in the basis states, where the entries of two basis states differ, and corresponds to the Hamming distance. For $|000\rangle_{\mathrm{L}}$, $\Delta n = 8$ and the decay factor is given by $e^{-4\gamma t}$. 
Global dephasing on the other hand will cause the off-diagonal elements to decay with a factor of $e^{-(\Delta m / 2)^2/2 \cdot \gamma t}$~\cite{pal2022relaxation}. $ \Delta m$ is the difference in magnetization of the basis states, where the magnetization of a state is given by the difference between the number of qubits in the ground state $|0\rangle$ and the remaining number of bits in the excited state
$|1\rangle$~\cite{pal2022relaxation}. 
\onecolumngrid
\twocolumngrid
For $|000\rangle_{\mathrm{L}}$, $\Delta m = 16$ and, thus, this prefactor is given by $e^{-32\gamma t}$. This means that, for $|000\rangle_{\mathrm{L}}$, the off-diagonal elements decay eight times faster for global dephasing than for local. $|000\rangle_{\mathrm{L}}$ is most sensitive to this global effect, since it is an eight-qubit GHZ-state with maximal difference in the magnetization between its basis states. This effect is expected and found to be less pronounced for $|+$$00\rangle_{\mathrm{L}} = \frac{1}{2} (|00000000\rangle + |11111111\rangle + |11001100\rangle + |00110011\rangle)$, where only some coherences decay according to $\Delta m = 8$ and some with $\Delta m = 4$, so twice and eight times faster than for local dephasing. We only account for local dephasing in our simulations, as characterized in App.~\ref{app:numerical_methods}, which may partly explain the observed differences between the numerically determined and the experimental fidelities.

\section{Error budget}\label{app:error_budget}
\begin{figure*}[!tb]
    \centering
    \includegraphics[width=180mm]{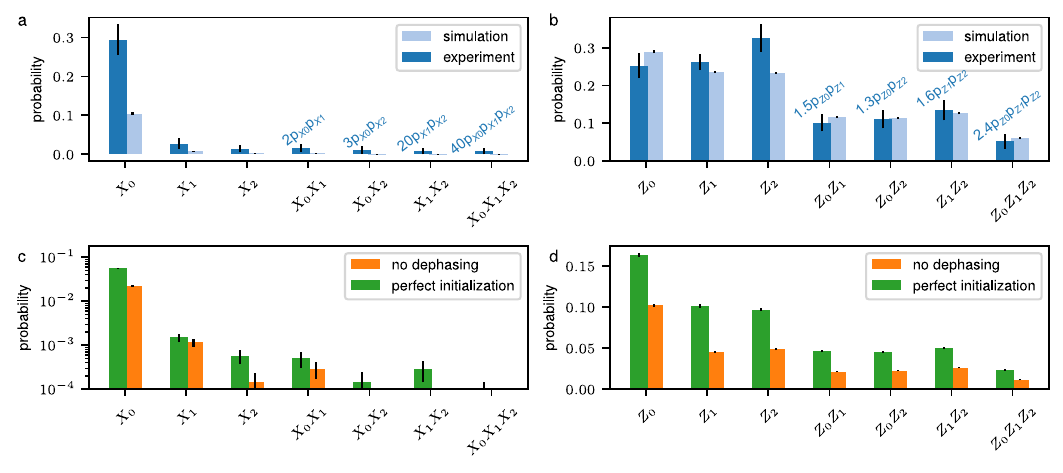}
    \caption{\justifying \textbf{Error budget for $H_\mathrm{L}$-gate injection. } \textbf{a, b,} Simulated and experimentally obtained probabilities for logical $X$- and $Z$-error configurations. The numbers on top of the bars correspond to the ratio between $p(X_{\mathrm{L}}^i X_{\mathrm{L}}^j)$ and $p(X_{\mathrm{L}}^i) p(X_{\mathrm{L}}^j)$, and $p(Z_{\mathrm{L}}^i Z_{\mathrm{L}}^j)$ and $p(Z_{\mathrm{L}}^i) p(Z_{\mathrm{L}}^j)$. For example the determined probability $p(X_{\mathrm{L}}^0 X_{\mathrm{L}}^1)$ is $2$ times larger than the probability one would expect from independent errors on logical qubits 0 and 1, $p(X_{\mathrm{L}}^0) p(X_{\mathrm{L}}^1)$. \textbf{c, d,} Probabilities for each logical $X$- and $Z$-error configuration without dephasing, and for a perfectly initialized input state. Here, the probabilities for $X$-error configurations are shown on a logarithmic scale for visibility. }
    \label{fig:error_budget}
\end{figure*}

Logical errors can be correlated in quantum codes that encode multiple logical qubits, such as block codes~\cite{poulin2006optimal} or quantum low-density parity-check codes~\cite{gottesman2013fault, breuckmann2021quantum}. 
We investigate these correlated errors by determining the probabilities for each logical error configuration, including single and correlated errors, for the non-transversal single-logical $H_{\mathrm{L}}$-gate on the $[[8, 3, 2]]$-code. To this end, we prepare logical state $|+$$00\rangle_{\mathrm{L}}$($|0+$$+\rangle_{\mathrm{L}}$), then apply the $H_{\mathrm{L}}$-gate to the first qubit and measure destructively in the $Z$($X$)-basis. From this, we infer if one, two or all three logical qubits have been flipped, which corresponds to the probability of logical $X$($Z$)-errors. 
Fig.~\ref{fig:error_budget}\textbf{a} and \textbf{b} show the probabilities for logical $X$- and $Z$-error configurations on the experimental setup. Notably, logical errors do not occur independently as $p(X_{\mathrm{L}}^i X_{\mathrm{L}}^j) \neq p(X_{\mathrm{L}}^i) p(X_{\mathrm{L}}^j)$ and $p(Z_{\mathrm{L}}^i Z_{\mathrm{L}}^j) \neq p(Z_{\mathrm{L}}^i) p(Z_{\mathrm{L}}^j)$, as theoretically predicted in previous works on quantum LDPC codes~\cite{old2024lift}. 
Fig.~\ref{fig:error_budget}\textbf{c} and \textbf{d} show numerical data for a setting without dephasing on idling qubits and for perfectly initialized logical states, to isolate the contribution of the $H_\mathrm{L}$-gate protocol. Logical error probabilities decrease substantially, while the overall distribution is maintained. Notably, we find that dephasing attributes for a large part of the overall logical error rate: without dephasing, the logical $Z$-error rate on qubit $0$ drops from almost $0.3$ (left most light blue column in Fig.~\ref{fig:error_budget}\textbf{b}) to less than $0.1$ (left most orange column in Fig.~\ref{fig:error_budget}\textbf{d}).

\section{Grover's search algorithm}\label{app:grover_search}

The number of required Grover iterations $n$ providing the highest amplification of the solution-states depends on the size of the search space $N$ ($N=2^3=8$ in our case) and the number of solutions $s$.
In this work, we use a phase oracle \cite{figgatt2017complete, qiskit_textbook} with two solutions ($s=2$) 
$w \in \{\ket{011}, \ket{101}\}$: states
$\ket{011}$ and $\ket{101}$ are marked by the oracle of the form
\begin{equation}
    O = \mathrm{C_1}Z_2 \cdot \mathrm{C_0} Z_2.
\end{equation}

The initial equal-superposition state can be represented as a superposition of solutions and non-solution states~\cite{Nielsen_and_Chuang}
\begin{align}
    \ket{\psi} &= \dfrac{1}{\sqrt{8}} \sum\limits_{\psi'' \notin \{011, 101\}} \ket{\psi''} + \dfrac{1}{\sqrt{8}}(\ket{011} + \ket{101}) \\&= \sqrt{\frac{N-s}{N}} \ket{\psi'} + \sqrt{\frac{s}{N}}\ket{w} = \cos \theta \ket{\psi'}+\sin \theta \ket{w}\nonumber
\end{align}
with $\sqrt{s/N}=\sin \theta$, i.e. $\theta = \pi/6$ in our case.
The probability of obtaining a valid solution $w$ when measuring in the computational basis is $s/N=1/4$ and the probability of obtaining an orthogonal non-solution state $\psi'$ equals to $(N-s)/N = 3/4$. 
One Grover step is a product of two reflections, first about the solution states $\ket{w}$ with the oracle $O$ and then about the initial state $\ket{\psi}$ with the diffusion operator $D$. This corresponds to an overall rotation of the initial state, and the rotation angle can be identified to be $2 \theta = \pi/3$ in our case. 
A single application of the Grover iteration including the oracle $O$ and the diffusion operator $D$, amplifies the probability of success to 1~\cite{Nielsen_and_Chuang, qiskit_textbook}, since
\begin{equation}
    D \cdot O \ket{\psi} = \cos ((2+1)\theta) \cdot \ket{\psi'} + \sin ((2+1)\theta )\cdot \ket{w} = \ket{w},
\end{equation}
meaning that a solution in the fault-free case is found with certainty. 
Analogously, the probability to find a solution after $k$ Grover iterations in a noise-free setting is given by $\sin^2( (2k+1) \theta)$.

A quantum circuit implementing this algorithm is shown in Fig.~\ref{fig:grover}\textbf{a}. This original circuit can be simplified to allow for a more simple implementation with logical qubits compiled into available logical gates of the $[[8, 3, 2]]$-code, reducing the number of required $H_\mathrm{L}$-gates to one. All operations in the resulting circuit can be fault-tolerantly implemented within the $[[8,3,2]]$-code as described in Sec.~\ref{sec:832_code}.

\section{Projected performance of Grover's algorithm}\label{app:projected_performance}

\begin{figure}[!tb]
    \centering
    \includegraphics[width=90mm]{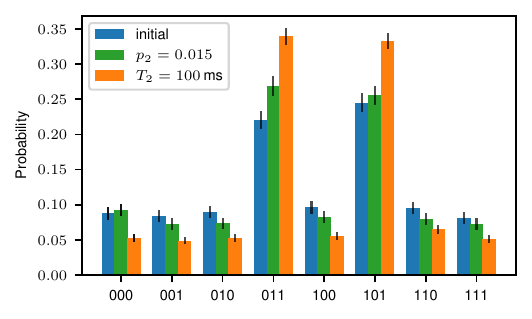}
    \caption{\justifying \textbf{Anticipated performance of the two-solution Grover search on logical qubits.} We simulate Grover's search algorithm for the set of noise parameters as characterized by the current experimental setup (blue), indicating a success probability of $p_{\mathrm{success}} = 0.40(4)$. For a slightly lower two-qubit-gate error rate of $p_2' = p_2 - 0.01 = 0.015$, we already obtain a total success probability of $0.52(1)$, which is above the classical optimal success probability of 0.46, as discussed in Sec.~\ref{sec:grover_search}. If instead of lowering $p_2$, we increase $T_2$ by a factor of 2 to \SI{100}{\milli\second} (orange), we find even higher success rates of $p_{\mathrm{success}} = 0.67(1)$. }
    \label{fig:projected_performance_grover}
\end{figure}

We simulate Grover's algorithm on logical qubits for different sets of noise parameters in order to estimate how much physical error rates have to improve to gain an advantage over the classically optimal success probability of 0.46. Fig.~\ref{fig:projected_performance_grover} shows the simulated probabilities to find each possible solution state for the initial set of noise parameters (blue), for a two-qubit-gate error rates reduced by a factor of 2 and for an increased coherence time $T_2' = 2 T_2 = \SI{100}{\milli\second}$. Both projected scenarios outperform the classical counterpart, indicating that even minor enhancements to the current setup could push performance beyond this break-even point. 

\section{Grover search on physical qubits}\label{app:physical_grover}

\begin{figure}[!tb]
    \centering
    \includegraphics[width=90mm]{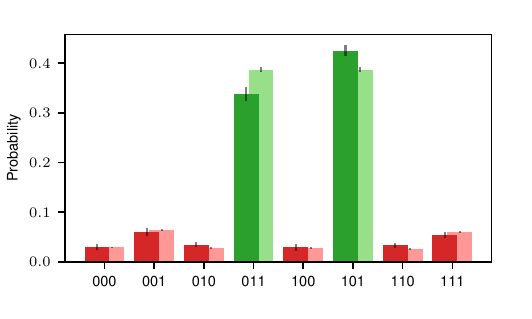}
    \caption{\justifying \textbf{Probabilities for the two-solution Grover search on physical qubits.} We implement the circuit shown in Fig.~\ref{fig:grover}\textbf{a} on physical qubits and determine the probabilities for each outcome in the experiment (darker columns) and simulation (lighter columns). }
    \label{fig:phycial_grover}
\end{figure}

We implement Grover's search algorithm on physical qubits, as compiled in Fig.~\ref{fig:grover}\textbf{a} on our experimental trapped-ion setup, accompanied by numerical simulations; the results are shown in Fig.~\ref{fig:phycial_grover}. The total experimental[simulated] success probability of 76(2)\%[77(1)\%] is larger than for the FT implementation on logical qubits, indicating that the quantum algorithm executed on logical qubits is currently still operated above the break-even point with its counterpart realization on physical qubits.

\clearpage
\bibliography{references}

\end{document}